\documentclass[doublecol]{epl2} 

\title{Non-equilibrium attractor for non-linear stochastic dynamics}

\author{A. Patrón\inst{1} \and B. Sánchez-Rey\inst{2} \and E. Trizac\inst{3,4} \and A. Prados\inst{1}}
\shortauthor{A. Patrón \etal}

\institute{                    
  \inst{1} Física Teórica, Universidad de Sevilla, Apartado de
  Correos 1065, E-41080 Sevilla, Spain\\
  \inst{2} Departamento de Física Aplicada I, E.P.S., Universidad de
  Sevilla, Virgen de África 7, E-41011 Sevilla, Spain\\
  \inst{3} LPTMS, Universit\'e Paris-Saclay, CNRS, 91405, Orsay, France\\
  \inst{4} Ecole Normale Sup\'erieure de Lyon, F-69364 Lyon, France
}

\usepackage{amsmath,amssymb}
\usepackage{physics}
\usepackage{cancel}
\usepackage{xcolor}

\newcommand{\LLNES}{\text{L}}
\newcommand{\eq}{\text{eq}}
\newcommand{\ther}{\text{th}}
\newcommand{\kin}{\text{kin}}
\newcommand{\eref}[1]{(\ref{#1})}

\abstract{
We study the dynamical behaviour of mesoscopic systems in contact with a thermal bath, described either via {a non-linear Langevin equation at the trajectory level---or the corresponding Fokker-Planck equation for the probability distribution function at the ensemble level.} Our focus is put on one-dimensional---or $d$-dimensional isotropic---systems in confining potentials, with detailed balance---fluctuation-dissipation thus holds, and the stationary probability distribution has the canonical form at the bath temperature. When quenching the bath temperature to low enough values, a far-from-equilibrium state emerges that rules the dynamics over a characteristic intermediate timescale. Such a long-lived state has a Dirac-delta probability distribution function and attracts all solutions over this intermediate timescale, in which the initial conditions are immaterial while the influence of the bath is still negligible. Numerical evidence and qualitative physical arguments suggest that the above picture extends to higher-dimensional systems, with anisotropy and interactions.
}

\begin{document}

\maketitle

Stochastic processes are ubiquitous in physics. Systems of interest are usually not isolated but in contact with a much larger environment. What makes their dynamics stochastic is the interaction with the environment (thermal bath): the integration over its degrees of freedom entails that the ``force''---understood in a generalised sense---acting on the system becomes effectively random~\cite{zwanzig_nonlinear_1973,di_cairano_derivation_2022}. It is in this approach, often called mesoscopic, that the Langevin equation emerges---see Ref.~\cite{schilling_coarse-grained_2022} for a recent review.

More than a century ago, Langevin initiated the approach that bears his name, when studying Brownian motion~\cite{langevin_sur_1908}.  This  is still an active field of research today: current experimental techniques make it possible to confine the Brownian particles in a potential, the profile of which can be controlled~\cite{ciliberto_experiments_2017,martinez_colloidal_2017,guery-odelin_driving_2023}. In turn, shaping the potential makes it possible to control the dynamical evolution, allowing for optimising observables such as irreversible work~\cite{aurell_optimal_2011,zhang_work_2020} or escape times~\cite{chupeau_optimizing_2020}, building smooth protocols that connect arbitrary states~\cite{plata_taming_2021}, or precisely designing finite-time computations~\cite{boyd_shortcuts_2022}.

The relevance of the Langevin approach is not restricted to Brownian motion; it is employed in a wealth of physical contexts, in which the above general picture for stochastic dynamics applies. Examples abound, including astrophysics~\cite{shalchi_analytical_2006,satin_classical_2018}, polymers~\cite{iyer_langevin_2018}, laser-cooled atoms~\cite{kindermann_nonergodic_2017,falasco_generalized_2022}, particle physics~\cite{ke_linearized_2018,capellino_fluid-dynamic_2022}, systems with negative temperatures~\cite{baldovin_langevin_2018,baldovin_derivation_2019}, {or optical spectroscopy~\cite{zhang_multidimensional_2023},} to name just a few. Interestingly, the analysis of experimental ``noisy'' data makes it possible to infer the underlying stochastic, Langevin-like, dynamical equations, not only in physics but also in neuroscience or biology~\cite{friedrich_approaching_2011,ferretti_building_2020,baldovin_langevin_2019,frishman_learning_2020,genkin_learning_2021}. Besides, since the early days of quantitative economy, related approaches making use of random walks are employed \cite{bachelier_theorie_1900,bouchaud_theory_2003}.

In the long-time limit, systems evolving under stochastic dynamics typically relax to equilibrium at the bath temperature. The equilibrium state is thus a global attractor, reached from an arbitrary initial condition, of the system dynamics~\cite{van_kampen_stochastic_1992}. A relevant question {is the following: before equilibrium, does the system reach a global non-equilibrium attractor, already independent of the initial preparation? In that case,} relaxation to equilibrium would proceed in two stages: first, the system would approach the universal non-equilibrium state and, second, this non-equilibrium state would tend to {the equilibrium one.}

In this Letter, we show---under general assumptions---that there emerges such a universal non-equilibrium state for a wide class of systems {in contact with a thermal bath,} when quenched to low enough temperatures. {Their dynamics is assumed to be Markovian and} described by a non-linear Langevin equation. This state, which we term long-lived non-equilibrium state (LLNES), is a global attractor of the dynamics for an intermediate time scale, over which initial conditions are immaterial but the system is still far from equilibrium. In particular, the probability distribution function (pdf) features a Dirac-delta shape within the LLNES. 

For the sake of concreteness, we focus here on the physical, intuitive ideas, that are behind the emergence of the LLNES in one-dimensional---or $d$-dimensional isotropic---systems; a more formal, mathematical, approach is presented in the Supplemental Material. Therein, we also provide numerical evidence on the existence of the LLNES for a more general situation, $d$-dimensional confining potentials---including anisotropy and interactions.

Let us now consider {a physical system with mesoscopic state  described by $\bm{r}\equiv\{x_{1},\ldots,x_{d}\}$.} A prototypical example is a colloidal particle confined in a $d$-dimensional
potential well. We assume the dynamics of $\bm{r}$ to be Markovian and governed by the following Fokker-Planck equation for the pdf $P = P(\bm{r},t)$,
\begin{align}
  \partial_{t} P=&\nabla_{\bm{r}}\!\cdot\!
  \left[ \bm{A}(\bm{r})P
  +\frac{1}{2} B^{2}(\bm{r})\nabla_{\bm{r}}P\right].
  \label{eq:FP}
\end{align}
{We stress the fact} that, in general, not only the ``force'' $\bm{A}(\bm{r})$ but also the diffusivity $B^2(\bm{r})$ are non-linear functions of $\bm{r}$. 
{Sometimes, this Fokker-Planck equation with non-linear coefficients is termed non-linear, although the equation is indeed linear in the probability distribution $P$~\cite{van_kampen_stochastic_1992}. This should be distinguished from other situations, where Fokker-Planck-like equations that are non-linear in the probability distribution are employed---see for instance~\cite{shiino_h-theorem_1985,shiino_dynamical_1987,bonilla_h-theorem_1996,frank_nonlinear_2005}}.

The dynamics of the system is stochastic due to its contact with a thermal bath at temperature $T$. We assume that detailed balance holds~\cite{horowitz_exact_2009,van_kampen_stochastic_1992}, so  the fluctuation-dissipation relation
\begin{equation}\label{eq:fluct-dissip}
  2\bm{A}(\bm{r})=\beta \, B^{2}(\bm{r})\,\nabla H(\bm{r}).
\end{equation} 
is verified, {with  $\beta=(k_{B}T)^{-1}$ and} $H(\bm{r})$ being the system's ``Hamiltonian''. In certain contexts, $H(\bm{r})$ would not be the Hamiltonian of the system but the function playing its role: e.g., for an overdamped Brownian particle, $H(\bm{r})$ would be the confining potential. Therefore, the canonical distribution, proportional to  $e^{-\beta H(\bm{r})}$, is the stationary solution of the Fokker-Planck equation \cite{gardiner_stochastic_2009}.

The Markov process $\bm{r}(t)$ can also be characterised by the Langevin equation at the trajectory level of description. When $B$ depends on $\bm{r}$, the noise is said to be ``multiplicative''~\cite{van_kampen_stochastic_1992} {and several Langevin formulations correspond to the same Fokker-Planck equation,}
\begin{equation}\label{eq:Langevin}
  \dot{\bm{r}}(t)=-\left[\bm{A}(\bm{r})-(\alpha-1) B(\bm{r})\nabla B(\bm{r})
  \right]+B(\bm{r})\bm{\eta}(t).
\end{equation}
Here, $\bm{\eta(t)}$ is the unit Gaussian white noise, $\expval{\eta_{i}(t)}=0$, $\expval{\eta_{i}(t)\eta_{j}(t')}=\delta_{ij}\delta(t-t')$ {and the ``multiplicative-noise'' parameter $\alpha$ must be chosen in the interval $[0,1]$~\cite{gardiner_stochastic_2009}. For each physical situation, the correct interpretation---typical ones are $\alpha=0$ for Ito's, $\alpha=1/2$ for Stratonovich's, $\alpha=1$ for Klimontovich's---of the Langevin equation with multiplicative noise is dictated by physics, not by mathematics~\cite{van_kampen_ito_1981,mannella_ito_2012}. If $B$ is constant, i.e., if the  noise is additive, $\alpha$ becomes irrelevant.

{Now} we consider a quench to a low temperature: the system is initially prepared at equilibrium at temperature $T_{i}$, and put in contact with a thermal bath at a much lower temperature $T_{f}$. In the subsequent relaxation to equilibrium at temperature $T_{f}$, there is a time regime in which noise is negligible: since $H$ is independent of the temperature, fluctuation-dissipation~\eref{eq:fluct-dissip} entails that $B^2(\bm{r})/|A(\bm{r})|\propto T_{f}\ll T_{i}$.} Therefore, terms containing $B(\bm{r})$ in Eq.~\eref{eq:Langevin} can be neglected and the Langevin equation reduces to the deterministic, noiseless equation
\begin{equation}\label{eq:noiseless-evol}
  \dot{\bm{r}}=-\bm{A}(\bm{r}),
\end{equation}
{which is independent of the parameter $\alpha$ in Eq.~\eref{eq:Langevin}.
}

{In what follows, we establish the conditions under which, for long enough times, the initial conditions are forgotten for the solution of Eq.~\eref{eq:noiseless-evol}. To be concrete, a simple but physically relevant situation with radial symmetry, $\bm{A}(\bm{r})=A(r)\bm{\hat{r}}$, 
$r=|\bm{r}|$, $\bm{\hat{r}}=\bm{r}/r$, is considered. The deterministic ``force'' $\bm{A}$ must be confining but otherwise arbitrary. This is indeed the case of the prototypical situation of a Brownian particle confined in an isotropic potential $U$, for which the Langevin equation reads
\begin{equation}\label{eq:LE-overdamped}
  \dot{\bm{r}}=-\gamma^{-1}\,U'(r)\hat{\bm{r}}+\sqrt{2D}\,\bm{\eta}(t),
\end{equation}
where $\gamma$ and $D$ are the friction and diffusion coefficients,
assumed to be position independent. The identifications $H=U$,
$\bm{A}=\gamma^{-1}U'(r)\bm{\hat{r}}$ and $B=\sqrt{2D}$ (thus additive noise) in the general
fluctuation-dissipation relation~\eref{eq:fluct-dissip} lead to the
Einstein relation $\beta\gamma D=1$. {Still, this is not the only physical situation, e.g. one may also address the relaxation of the velocity of a colloidal particle due to the non-linear drag force stemming from its interaction with the background fluid, considered later. Therein, the variable $\bm{r}$ would stand for the velocity of the particle.} Note that, since $A$ may change sign as $r$ decreases, the potential may have several minima. 

{
From Eq.~\eref{eq:noiseless-evol}, the time evolution for one trajectory starting from $r_i$ is implicitly given by
\begin{equation}\label{eq:x(t)-implicit}
  t=\int_{r(t)}^{r_i}\frac{dr'}{A(r')}, \quad r_i\equiv r(t=0).
\end{equation}
Assuming that 
\begin{equation}\label{A(x)-small-large-x}
  \quad \lim_{r\to+\infty}r^{-1}A(r)=+\infty, 
\end{equation}
i.e. $A$ diverging faster than linearly for large $r$, we have
\begin{equation}\label{eq:x(t)-implicit-asymp}
t =
\int_{r(t)}^{+\infty}\frac{dr'}{A(r')}-\int_{r_i}^{+\infty}\frac{dr'}{A(r')},
\end{equation}
when the confining is stronger than harmonic at large distances. The first (second) term  on the rhs of Eq.~\eref{eq:x(t)-implicit-asymp} is the time needed to relax from a very large value of $r$, much larger than $r_i$, to the instantaneous position $r(t)$ ($r_i$).}

{
Let us assume that the initial temperature $T_i$ is much larger than the final one $T_f$, implying the following timescale separation
\begin{equation}\label{eq:LLNES-interm-times}
  t_{1}\equiv\tau(T_{i})\ll t \ll t_{2}\equiv\tau(T_{f}),
\end{equation}
where $\tau(T)$ is the relaxation time to equilibrium at temperature $T$. In this way, there appears an intermediate time regime, in which the second term on the rhs of Eq.~\eref{eq:x(t)-implicit-asymp} is negligible against the first {while} noise is still irrelevant.  Over the timescale in Eq.~\eref{eq:LLNES-interm-times}, we thus get
}
\begin{equation}\label{eq:LLNES-def}
  r(t)\sim r_{\LLNES}(t), \quad  \int_{r_{\LLNES}(t)}^{+\infty} \frac{dr}{A(r)}=t.
\end{equation} 
The state $r_{\LLNES}(t)$ defined in Eq.~\eref{eq:LLNES-def} is a non-equilibrium attractor of the dynamics of the system, {all solutions of the Langevin equation tend to it over the timescale defined in eq.~\eref{eq:LLNES-interm-times}, independently of their initial condition}. We term $r_{\LLNES}(t)$ long-lived non-equilibrium state (LLNES).\footnote{This terminology was already employed in Ref.~\cite{patron_strong_2021} for a specific form of $A(\bm{r})$ in the context of non-linear Brownian motion.} {Note that $t_1$ and $t_2$ are thus determined by the conditions $r_{\LLNES}(t_1)=r_i$ and $r_{\LLNES}(t_2)=r_f$, respectively.
}

Over this far-from-equilibrium state, independent of  initial conditions, the pdf is
\begin{equation}\label{eq:P-LLNES-x}
  P_{\LLNES}(r,t)\sim \delta (r-r_{\LLNES}(t)), 
\end{equation}
{as formally shown in the Supplemental Material.}
Throughout the paper, we use the symbol $\sim$ with the meaning of ``asymptotic to''~\cite{bender_advanced_1999}, i.e. $f(x)\sim g(x)$ for $x\to x_0$ means that $\lim_{x\to x_0}f(x)/g(x)=1$. The function $r_{\LLNES}(t)$ defined by Eq.~\eref{eq:LLNES-def} depends on the specific form of the function $A(r)$. However, we can introduce a scaled variable $c$ such that its corresponding pdf is universal and time-independent,
\begin{equation}\label{eq:P-LLNES-xi}
  \bm{c}\equiv {\bm{r}}/{\expval{r(t)}}, \qquad
  P_{\LLNES}(c,\cancel{t})\sim \delta (c-1). 
\end{equation}
We recall that, over the LLNES, $\expval{r(t)}=r_{\LLNES}(t)$. {Note that the terms containing $B(r)$ in the Langevin equation~\eref{eq:Langevin} eventually drive the system to equilibrium at $T_{f}$. In other words, the LLNES is ``destroyed'' for long enough times, when $r_{\LLNES}(t)=O(\expval{r}_{\eq}(T_{f}))$, i.e.~as $t=O(t_2)$.}

We now apply the results presented here to two different physical situations.
First, we consider the confined Brownian particle of Eq.~\eref{eq:LE-overdamped}, particularised for the non-linear potential
\begin{equation}
\label{eq:nonlinear-potential}
  U(r)=\frac{1}{2}k r^{2}+\frac{1}{4}\lambda r^{4}, \quad \lambda>0.
\end{equation}
The condition $\lambda>0$ ensures that the potential is confining: 
\begin{equation}
    A(r)=ar+br^{3}, \quad a\equiv k/\gamma, \; b\equiv \lambda/\gamma.
\end{equation} 
Moreover, Eq.~\eref{A(x)-small-large-x} holds and we have the necessary timescale separation. 

We analyse the case $k>0$ to start with, in which the ``force'' $A(r)>0$ $\forall r\ne 0$ and $U(r)$ has only one minimum at the origin. Later, we consider the case $k<0$, which corresponds to a ``lemon-squeezer'' potential with multiple minima at $r=r_c$, {where}
\begin{equation}
    r_c\equiv\sqrt{|a|/b}=\sqrt{|k|/\lambda}.
\end{equation}

For $k>0$, Eq.~\eref{eq:LE-overdamped} reduces to
\begin{equation}\label{eq:overdamped-langevin}
  \dot{\bm{r}}=-ar \left(1+r^2/r_c^2\right)\hat{\bm{r}}+\sqrt{2D}\,\bm{\eta}(t).
\end{equation}
In this physical situation, there are two characteristic lengths, 
\begin{equation}
    r_{\lambda}\equiv (k_B T/\lambda)^{1/4}, \quad r_{k}\equiv (k_B T/k)^{1/2},
\end{equation}
which---aside from constants---correspondingly give the equilibrium lengths at high and low temperatures. In fact, it is useful for our analysis to introduce a  dimensionless temperature 
{
\begin{equation}
    T^*=k_B T \lambda/k^2=(r_k/r_\lambda)^4,
\end{equation}
}
high and low temperatures thus correspond to the regimes $T^*\gg 1$ and $T^*\ll 1$, respectively. {Note that $r_c=r_{\lambda}^2/r_k$.}

Let us analyse the emergence of the LLNES in this specific situation. The particularisation of Eq.~\eref{eq:x(t)-implicit-asymp} gives
\begin{equation}\label{eq:explicit-sol-nonlin-pot}
  2at=
  \ln\left(1+{r_c^2}/{r^{2}(t)}\right)-\ln\left(1+{r_c^2}/{r_i^{2}}\right).
\end{equation}
{For} a high enough initial temperature $T_i^*\gg 1$, 
we estimate $r_i$ with $r_{\lambda,i}=(k_B T_i/\lambda)^{1/4}$. There appears an intermediate time window over which $r_i \gg r(t) \gg r_c$ {and} initial conditions are forgotten, specifically
\begin{equation}\label{eq:LLNES-explicit-short-times-v2}
  r(t)\sim r_{\LLNES}(t)=(2bt)^{-1/2}, \quad  (T_i^*)^{-1/2} \ll 2at\ll 1  .
\end{equation}
{Note that} $r_{\LLNES}(t)$ only depends on $b=\lambda/\gamma$, i.e.~only on the behaviour of the potential at large distances. 

{In order to derive Eq.~\eref{eq:LLNES-explicit-short-times-v2}, it is only necessary to consider a high enough initial temperature;} the role of the final temperature is to (possibly) limit the timescale over which the LLNES is observed. {Noise is negligible as long as  $r_{\LLNES}(t)$ is much larger than the equilibrium value at the final temperature, $r_{k,f}=(k_B T_f/k)^{1/2}$, which gives the condition $2at \ll (T_f^*)^{-1}$. If $T_f^*=O(1)$ or larger, this restricts the  LLNES in Eq.~\eref{eq:LLNES-explicit-short-times-v2} to the time window  $(T_i^*)^{-1/2}\ll 2at \ll (T_f^*)^{-1}$. If $T_f^*\ll 1$, the LLNES extends to longer times such that $2at=O(1)$, $r(t)$ becomes of the order of $r_c$ and}
\begin{equation}\label{eq:LLNES-explicit-specific}
  r_{\LLNES}(t)=r_c \left(e^{2at}-1\right)^{-1/2}.
\end{equation}
Figure~\ref{fig:trajectories} shows a set of stochastic trajectories for which the behaviours in Eqs.~\eref{eq:LLNES-explicit-short-times-v2} and \eref{eq:LLNES-explicit-specific} are observed.
\begin{figure}
\includegraphics[width=3.4in]{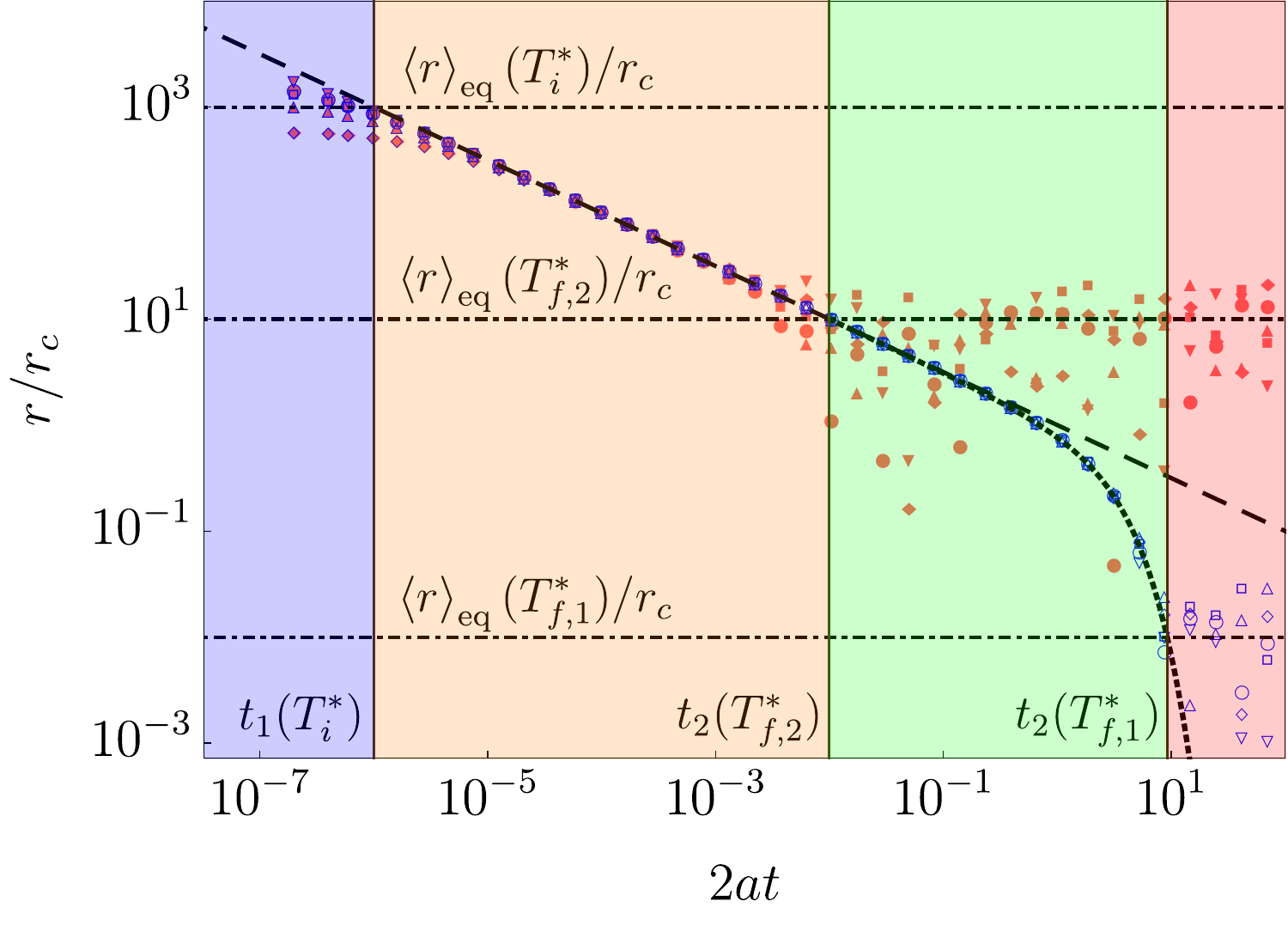}
\caption{Stochastic trajectories for the overdamped Brownian particle in a non-linear potential. {Five realisations of the Langevin Equation~\eref{eq:overdamped-langevin} for $d=1$, sampled from an initial equilibrium state such that
$\expval{r}_{\text{eq}}(T_i^*)/r_c=10^3$,  to two final states such that $\expval{r}_{\text{eq}}(T_{f,1}^*)/r_c=10^{-2}$ (blue empty symbols) and $\expval{r}_{\text{eq}}(T_{f,2}^*)/r_c=10$ (red filled symbols), are shown.}
Black vertical lines correspond to the characteristic times that delimit the different coloured regions: (1) $t <t_1(T_i^*)$ (blue), where initial conditions still prevail, (2) $t_1(T_i^*) <t<t_2(T_{f,2}^*)$ (orange), where the power law Eq.~\eref{eq:LLNES-explicit-short-times-v2} (dashed line) applies for both final temperatures, (3) $t_2(T_{f,2}^*)< t \ll t_2(T_{f,1}^*)$ (green), where Eq.~\eref{eq:LLNES-explicit-specific} (dotted curve) applies for $T_{f,1}^*$ whereas thermal noise becomes relevant for $T_{f,2}^*$, and (4) $t> t_2(T_{f,1}^*)$ (pink), where noise becomes relevant for $T_{f,1}^*$. 
}
    \label{fig:trajectories}
\end{figure}

We now study the case $k<0$, the ``lemon-squeezer'' potential with multiple minima at $r=r_c$. In the one-dimensional situation,  the potential would be bistable, with two symmetric minima. The LLNES in Eq.~\eref{eq:LLNES-explicit-short-times-v2}, which only depends on the details of the potential at large $r$, is still present for $T_i^*\gg 1$; it is thus independent of the presence of other minima. Also, the LLNES extends to longer times if $T_f^*\ll 1$, but it is no longer given by Eq.~\eref{eq:LLNES-explicit-specific}, we have 
\begin{equation}
    r_{\LLNES}(t)=r_c\left(1-e^{-2at}\right)^{-1/2}
\end{equation}
instead. The system reaches equilibrium at $r_c$ over this regime, with small thermal fluctuations{---see the Supplemental Material for a more detailed discussion.}
\begin{figure*}
  \centering
  {\centering \includegraphics[width=3in]{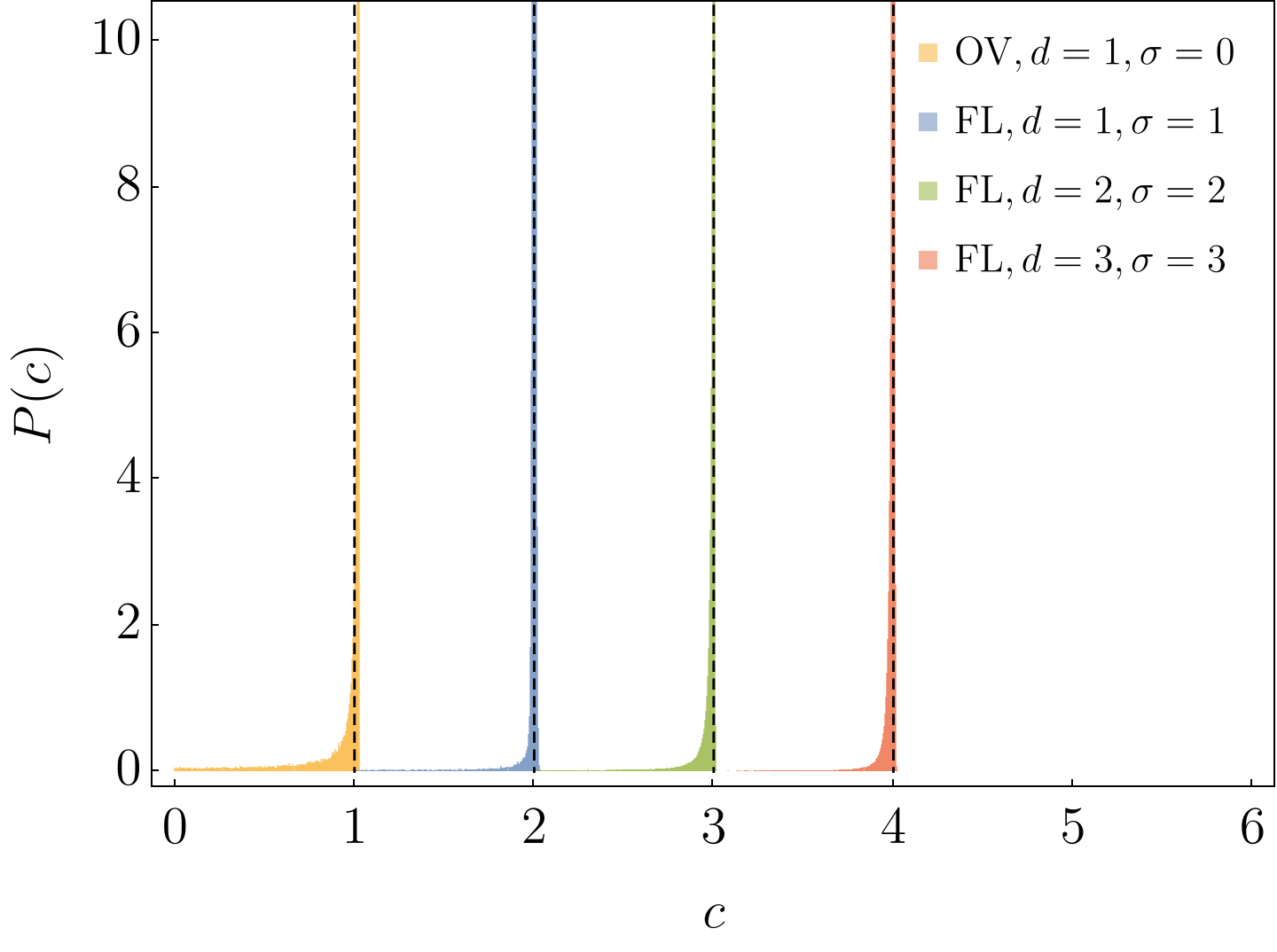}\hspace*{2em}
    \includegraphics[width=3.25in]{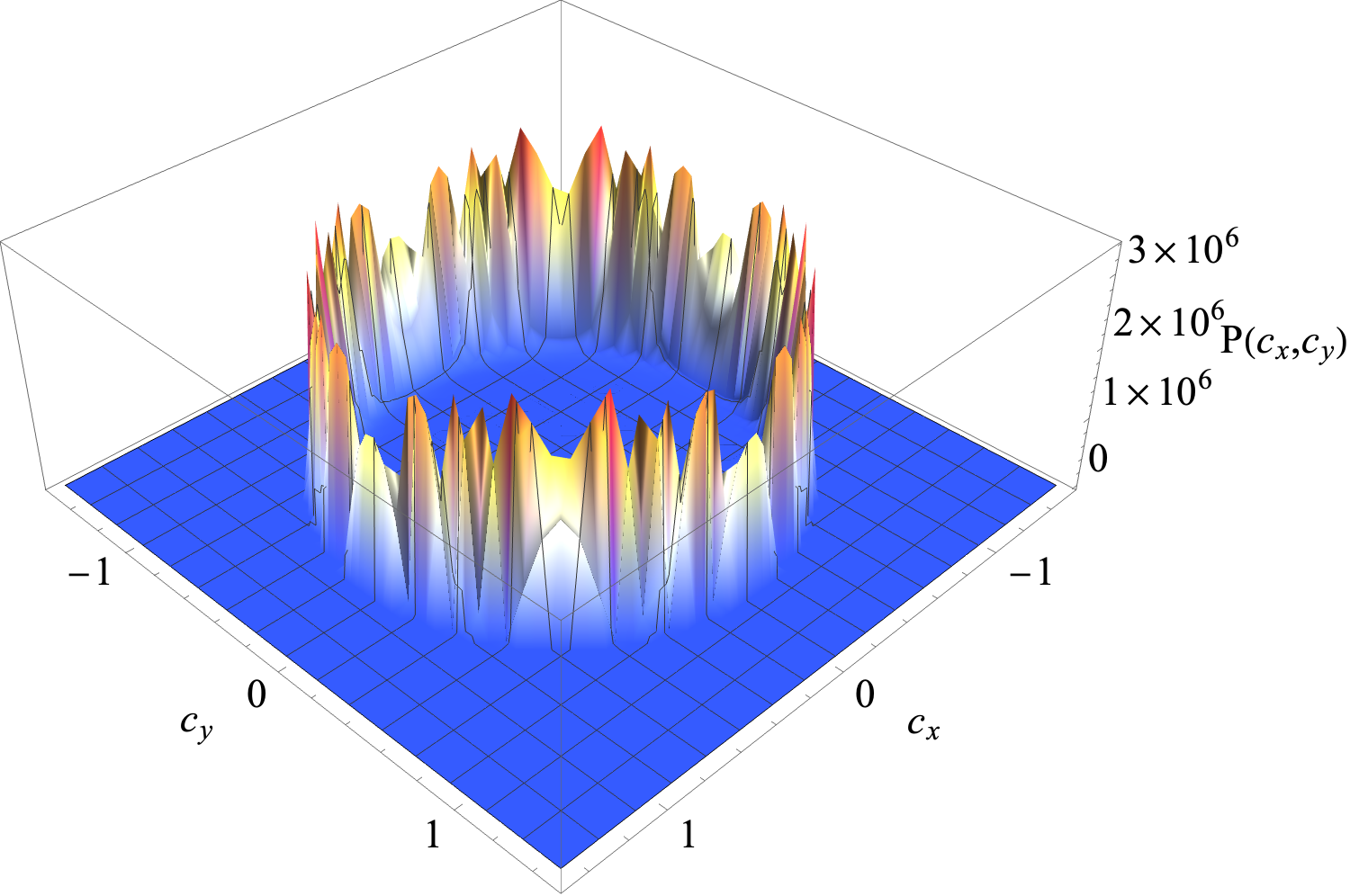} }
  \caption{Scaled pdf in the LLNES for different physical
    situations. (Left) Plot for both (OV) the overdamped particle in a non-harmonic potential, Eq.~\eref{eq:overdamped-langevin}, and (FL) the molecular fluid with non-linear drag, Eq.~\eref{eq:FP-fluid}, for different spatial dimensions. {For the former, $\bm{c}=\bm{r}/\expval{r}$; for the latter, $\bm{c}=\bm{v}/\expval{v}$.} In order to appreciate the universal Dirac-delta shape in Eq.~\eref{eq:P-LLNES-xi}, each pdf is shifted a factor $\sigma$ to the right, as indicated in the legend. In each case, a small tail to the left of the peak is observed{---see the Supplemental Material for a detailed account of the (small) deviations from the delta peak.} (Right) Plot of the bivariate scaled pdf in the LLNES for the molecular fluid in the two-dimensional case.}
  \label{fig:deltas}
\end{figure*}

Now we consider another relevant physical system: an isotropic fluid
with non-linear drag force. {Specifically, we investigate the stochastic evolution of $N$ particles undergoing binary collisions and immersed in a background fluid acting as a thermal bath. For dilute enough systems, the velocity pdf $P(\bm{v},t)$ obeys the Boltzmann-Fokker-Planck equation~\cite{santos_mpemba_2020,patron_strong_2021,megias_thermal_2022}}
\begin{equation}\label{eq:FP-fluid}
    \partial_t P=\nabla_{\bm{v}}\cdot\left[\zeta(v)\left(\bm{v}+\frac{k_B T}{m}\nabla_{\bm{v}}\right)P\right]+J[P,P],
\end{equation}
where $\zeta(v)$ stands for the velocity-dependent drag
coefficient {and $J[P,P]$ is the Boltzmann collision term, which is bilinear in $P${---see the Supplemental Material for more details.} For low velocities, the drag force is usually linear in $v$, $\lim_{v\to 0}\zeta(v)= \zeta_0$. For large velocities, the drag force may become non-linear in $v$: the dimensionless drag coefficient $\zeta^*\equiv\zeta/\zeta_0$ thus depends on $v$, as is the case {when} the masses of the Brownian and background fluid particles are comparable~\cite{ferrari_particles_2007,ferrari_particles_2014,hohmann_individual_2017}.} If collisions among particles are elastic, this system tends to the canonical distribution with $H(v) = mv^{2}/2$, provided that $A$ and $B$ are such that Eq.~\eref{eq:fluct-dissip} holds. Since $A(v) = \zeta(v)v$, we need $B^2(v)=2 \zeta(v) k_B T/m$; noise is thus multiplicative.

{The kinetic temperature is 
\begin{equation}
    T_{\kin}(t)\equiv \frac{m}{dk_B}\expval{v^2}(t),
\end{equation}
which equals the bath temperature at equilibrium. Initially, the system is equilibrated at $T_i$, thus $T_{\kin}(t=0)=T_i$, and the bath temperature is suddenly quenched to $T_f\ll T_i$. To be concrete, we restrict ourselves to drag coefficients with algebraic behaviour for large $v$, 
\begin{equation}
    \zeta^*(v)\sim\gamma (v/v_{\ther,f})^n, \quad v_{\ther,f}\equiv (2 k_B T_f/m)^{1/2},
\end{equation} 
with $\gamma$ being the non-linearity parameter and $v_{\ther,f}$ the thermal velocity at $T_f$. If $n>1$, there appears a timescale over which the non-linear drag dominates and both noise and collisions---even if they are inelastic---are  negligible. Over this wide time window, initial conditions are forgotten and the LLNES emerges. {Specifically, we have}
\begin{equation}
  v_{\LLNES}(t)/v_{\ther,f}=(\gamma \zeta_0n t)^{-1/n}, \quad (T_f/T_i)^{n/2}\ll n \gamma\zeta_0 t \ll 1,
\end{equation}
{as derived in the Supplemental Material. It is worth noting the} strong analogy with Eq.~\eref{eq:LLNES-explicit-short-times-v2}. The kinetic temperature thus shows a slow non-exponential, algebraic, decay as $T_{\kin}(t)\propto t^{-2/n}$,  which rules the emergence of memory effects such as the Kovacs and Mpemba effects~\cite{patron_strong_2021}.}

Figure~\ref{fig:deltas} shows the pdf of the scaled variable $\bm{c}$, 
for the two specific examples of physical systems described above.  The delta-peak structure is clearly observed, for one-, \mbox{two-,} and three-dimensional systems. {For the non-linear fluid, the data shown corresponds to $n=2$.}

In this Letter, we have analysed the dynamical behaviour of a wide class of physical systems, described by a non-linear Langevin (or {the corresponding} Fokker-Planck) equation {with detailed balance}. When quenched to a low enough temperature, all these systems reach a universal long-lived non-equilibrium state, regardless of initial conditions. This state, which we have termed LLNES, is characterised by a Dirac-delta pdf. 

{There} are two main hypotheses for the emergence of the LLNES: (i) the non-linearity of the ``force'' in the Langevin equation and (ii) the separation 
of the initial temperature $T_i$ from the final one $T_f$,
$T_i\gg T_f$. A separation of time scales ensues, with the 
LLNES appearing in the intermediate window, where 
initial conditions are irrelevant and noise is negligible.
Under these quite general assumptions, our results are independent of both the nature of the noise (either additive or multiplicative) and the dimensionality of the system, as shown in Fig.~\ref{fig:deltas}.

For the sake of simplicity, we have restricted the discussion to isotropic situations, in which our work proves the existence of the LLNES. {For the same reason, our terminology has been mainly focused on overdamped particles in confining potentials. Still, the presented framework is more general, since the physical meaning of the variables $x_i$ in the $d$-dimensional mesoscopic state $\bm{r}\equiv\{x_{1},\ldots,x_{d}\}$ and the ``force'' $\bm{A}(\bm{r})$ are not predetermined---as shown by our analysis of the fluid with non-linear drag, in which the $x_i$ stand for the components of the velocity. In a generic situation, the $x_i$ may include both positions and velocities, and thus our framework in principle applies to underdamped systems.  The Langevin equation for an underdamped particle of mass $m$ with non-linear drag force in a confining potential would be
\begin{equation}\label{eq:Langevin-underdamped}
    \dot{\bm{x}}=\bm{v}, \quad \dot{\bm{v}}=-\zeta(v) \bm{v}-\frac{1}{m} \nabla U(\bm{x})+ \sqrt{\frac{2\zeta(v) k_B T}{m}}\,\bm{\eta}.
\end{equation}
Note that this equation does not belong in the general class considered in Eq.~\eqref{eq:Langevin}. To include Eq.~\eqref{eq:Langevin-underdamped} in our framework, we have to generalise Eq.~\eqref{eq:Langevin} to the case in which $B$ is no longer a scalar, affecting all the components of $\bm{r}\equiv\{\bm{x},\bm{v}\}$ in the same way, but a second-rank tensor.\footnote{{In the simple case depicted in Eq.~\eqref{eq:Langevin-underdamped}, $B$ would be diagonal, with the elements corresponding to $\bm{x}$ equal to zero and the elements corresponding to $\bm{v}$ equal to $\sqrt{2\zeta(v) k_B T/m}$.}}  The non-linearity of the ``force'' $\bm{A}=\{-\bm{v},\zeta(v)\bm{v}+\nabla U(\bm{x})/m\}$ suggests that LLNES may also emerge in underdamped systems. Certainly, this constitutes an interesting perspective for future research.} 

It is always the form of the ``force'' at large distances that controls the emergence and shape of the LLNES, as illustrated by our analyses of the quartic potential and the non-linear fluid above. The effective reduction to one degree of freedom stemming from isotropy have allowed us to obtain analytical results for the emergence of the LLNES. The Supplemental Material presents formal proofs for the one degree of freedom case, and also numerical evidence and qualitative, physical, arguments that hint at the  the existence of the LLNES for more complex scenarios with several degrees of freedom---including anisotropy and interactions. {Note that the nature of the interactions is immaterial for our results---as long as the two main hypothesis (i) and (ii) above are fulfilled. Yet, either numerically checking the emergence of the LLNES or rigorously proving the conditions for its existence in these more complex situations, with anisotropy and interactions, are highly non-trivial tasks that lie beyond the goals of the present work.}

Quasi-elastic one-dimensional granular systems have been shown to display Dirac-delta pdfs~\cite{benedetto_kinetic_1997,barrat_velocity_2002,baldassarri_influence_2002} resembling that of the LLNES. This result was  derived from the inelastic Boltzmann equation, and therefore it cannot be considered as a particular case of the general result derived in this Letter---obtained within the Langevin framework. Still, the similarity of the observed pdfs entails it is worth investigating possible connections between these two intrinsically different physical situations. {On another note, the LLNES found here displays some similarities with the quasi-stationary states (QSS) observed in some systems with long-range interactions, such as the HMF model~\cite{latora_non-gaussian_2001,rapisarda_nonextensive_2005}. The possible existence of a deeper connection between the LLNES and these long-lived QSS is also worth investigating.}

{Finally,} testing the emergence of the LLNES in real experiments is an interesting prospect for future work. In particular, it seems worth exploring the relevance of the LLNES to control the time evolution of mesoscopic systems, {like biomolecules or memory devices.  In this regard, it must be stressed that the two specific examples considered here describe actual physical systems. Current techniques make it possible to control the shape of the potential confining a colloidal particle immersed in a fluid~\cite{ciliberto_experiments_2017,martinez_colloidal_2017}, and the Langevin equation for the velocity with non-linear drag has been successfully employed to describe mixtures of ultracold atoms~\cite{hohmann_individual_2017}.

\acknowledgments
    A. Patrón, B. Sánchez-Rey and A. Prados acknowledge financial support from Grant PID2021-122588NB-I00 funded by MCIN/AEI/10.13039/501100011033/ and by ``ERDF A way of making Europe''. All the authors acknowledge financial support from Grant ProyExcel\_00796 funded by Junta de Andalucía's PAIDI 2020 programme. A. Patr\'on acknowledges support from the FPU programme through Grant FPU2019-4110, and also additional support from the FPU programe through Grant EST22/00346, which funded his research stay at Univ.~Paris-Saclay during autumn 2022. A.~Prados also acknowledges the hospitality of LPTMS, which funded his stay at Univ.~Paris-Saclay in June 2022.

\bibliography{Mi-biblioteca-09-dic-2023}

\end{document}